\documentclass[12pt,dvips]{amsart}
\usepackage{euler, amsfonts, amssymb, latexsym, epsfig,epic,wasysym}

\theoremstyle{definition}

\theoremstyle{remark}

\numberwithin{equation}{section}
\newtheorem*{corollary*}{Corollary}
\newtheorem*{definition*}{Definition} 
\newtheorem*{theorem*}{Theorem}
\newtheorem*{conjecture*}{Conjecture}

\author{P. Di Francesco}
\address{P. Di Francesco,
Service de Physique Th\'eorique de Saclay,
CEA/DSM/SPhT, URA 2306 du CNRS,
C.E.A.-Saclay, F-91191 Gif sur Yvette Cedex, France}
\thanks{The authors acknowledge the support
of the European networks ``ENIGMA'' MRT-CT-2004-5652, ``ENRAGE'' MRTN-CT-2004-005616,
and of the Geocomp project (ACI Masse de Donn\'ees).}

\author{P. Zinn-Justin}
\address{P. Zinn-Justin,
Laboratoire de Physique Th\'eorique et Mod\`eles Statistiques, UMR 8626 du CNRS,
Universit\'e Paris-Sud, B\^atiment 100,  F-91405 Orsay Cedex, France}

\title{From Orbital Varieties to Alternating Sign Matrices}


\newcommand{\C}{\mathbb{C}}

\begin{document}
\pagestyle{plain}

\begin{abstract}
We study a one-parameter family of vector-valued polynomials associated to each simple Lie algebra.
When this parameter $q$ equals $-1$ one recovers Joseph polynomials, whereas at $q$ cubic root
of unity one obtains ground state eigenvectors of some integrable models with boundary conditions
depending on the Lie algebra; in particular, we find that the sum of its entries is related to 
numbers of Alternating Sign Matrices and/or Plane Partitions in various symmetry classes.


\end{abstract}

\maketitle  

\section{Introduction}
%
%
Recently, a remarkable connection between integrable models and combinatorics
has emerged. It first appeared in a series of papers concerning
the XXZ spin chain and the Temperley--Lieb (TL) loop model \cite{BdGN,RSa}
and which culminated
with the so-called Razumov--Stroganov (RS) conjecture \cite{RS}.
One of the main observations of \cite{BdGN}, a weak corollary of the RS conjecture,
is that the sum of entries of the properly normalized ground state vector of
the TL(1) loop model is (unexpectedly!) equal to the number of Alternating Sign
Matrices. This result was eventually proved in \cite{DFZJ} by using the {\it integrability}\/
of the TL loop model in the following way: the model is generalized by
introducing $N$ complex numbers (spectral parameters, or inhomogeneities)
in the problem, where $N$ is the size of the system. The ground state
entries become polynomials in these variables, and integrability provides many new tools for
analyzing them, leading eventually to the exact computation of their sum, identified as the so-called
Izergin--Korepin (IK) determinant, known to specialize
to the number of Alternating Sign Matrices in the homogeneous limit \cite{Kup}.
Note that in this work, the meaning of the spectral parameters is not very transparent;
in particular, it is unclear how to generalize the full RS 
conjecture in their presence.

Next, it was observed in \cite{Pas} that the polynomials obtained above really belong
to a one-parameter family of solutions of a certain set of linear equations, in which the parameter
$q$ has been set equal to a cubic root of unity. 
This observation is not obvious because
the equations for generic $q$ are not a simple eigenvector equation; in fact,
as explained in \cite{DFZJc}, they are precisely the quantum Knizhnik--Zamolodchikov ($q$KZ)
equations at level 1 for the algebra $U_q(\widehat{\mathfrak{sl}(2)})$.
Furthermore, in the ``rational'' limit $q\to -1$, these polynomials have a remarkable
geometric interpretation: they are equivariant Hilbert polynomials (or ``multidegrees'')
of $A_{N-1}$ orbital varieties $M^2=0$ (\cite{DFZJc}, see also \cite{KZJ}), which are extensions
of the Joseph polynomials \cite{Jo}. 
Note that here, the spectral parameters quite naturally appear
as the basis of weights of $\mathfrak{gl}(N)$. In \cite{DFZJc}, these ideas were generalized
to higher algebras $U_q(\widehat{\mathfrak{sl}(k)})$, which correspond to
the orbital varieties $M^k=0$.

Here, we pursue a {\it different}\/ type of generalization: 
we investigate orbital varieties corresponding to the other infinite series
of simple Lie algebras: $B_r$, $C_r$, $D_r$;
but we stick to the $U_q(\widehat{\mathfrak{sl}(2)})$ case
by choosing the orbital varieties $M^2=0$, $M$ a complex matrix 
in the fundamental representation.
Indeed, we show below that such orbital varieties are related to the
same loop model, but with different {\it boundary conditions} (corresponding
to variants of the Temperley--Lieb algebra).
Furthermore, one can now $q$-deform the resulting polynomials 
to produce solutions of $q$KZ equations of type $B$, $C$, $D$
and set $q$ to be
a cubic root of unity. Taking the homogeneous limit, the entries become
integer numbers, which we conjecture to be related to 
{\it symmetry classes}\/ of Alternating
Sign Matrices and/or Plane Partitions; in particular we identify the sums of entries.

In what follows we state most results without proofs; some will appear in
a joint paper with A.~Knutson \cite{DFKZJ} on a closely related subject.

\section{General setup}
\subsection{Orbital varieties}
Let $\mathfrak{g}$ be a simple complex Lie algebra of rank $r$, $\mathfrak{b}$
a Borel subalgebra. $\mathfrak{b}=\mathfrak{t}\oplus \mathfrak{n}$ where 
$\mathfrak{t}$ is the corresponding
Cartan subalgebra and $\mathfrak{n}$ is the space of nilpotent elements of $\mathfrak{b}$.
$B$ and $T$ are Borel and Cartan subgroups.
Let $W$ denote the Weyl group of $\mathfrak{g}$, and $s_\alpha$
its standard generators, where $\alpha$ runs over the set 
of simple roots of $\mathfrak{g}$.

Fixing an orbit $G\cdot x$, with $x \in \mathfrak{n}$ and $G$ acting by conjugation,
one can consider the irreducible components of
$\overline{{\mathfrak b}\cap (G\cdot x)}$, which are called orbital varieties.

Even though much of what follows can be done for any orbital varieties,
we focus below on the following special case: we fix an irreducible representation
$\rho$ (of dimension $N$) and consider 
the scheme $E=\{ x\in \mathfrak{b} \mid \rho(x)^2=0 \}$. The underlying
set is precisely a $\overline{{\mathfrak b}\cap (G\cdot x)}$, where $x$ is any element of $E$
such that $\rho(x)$ is of maximal rank.
In some sense, its components  are the ``simplest possible'' orbital varieties.

\subsection{Hotta construction}
It is known that there exists a representation of the Weyl group $W$ on the vector space
$V$ of formal linear combinations
of orbital varieties (Springer/Joseph representation); for each $G$-orbit, it is an irreducible
representation. We use the following explicit form of the representation:
note that orbital varieties are invariant under $T\times \C^\times$, where $T$ 
acts by conjugation and $\C^\times$ acts by overall scaling. We can therefore consider equivariant
cohomology $H^*_{T\times \C^\times}(\cdot)$ and in particular via 
the inclusion map from each orbital
variety $\pi$ to the space $\mathfrak{n}$,
the unit of $H^*_{T\times\C^\times}(\pi)$ is pushed forward
to some cohomology class $\Psi_\pi$ in $H^*_{T\times\C^\times}(\mathfrak{n})=\C[\mathfrak{t},A]$,
that is a polynomial in $r+1$ variables $\alpha_1$, $\ldots$, $\alpha_r$, $A$ (the $r$ simple
roots plus the $\C^\times$ weight), sometimes
called {\it multidegree}\/ of $\pi$. Suppressing the $\C^\times$ action, that is setting $A=0$,
one recovers the Joseph polynomials \cite{Jo}.

The way that $W$ acts on these polynomials can be described explicitly, by extending slightly the
results of Hotta \cite{Ho} to include the additional $\C^\times$ action. 
One starts by associating to each simple root $\alpha$ a certain geometric construction, which we briefly
recall. For $x\in \mathfrak{b}$ write $x=\sum_\alpha x_\alpha e_\alpha$ where $\alpha$ runs over
positive roots, $e_\alpha\in\mathfrak{g}$ being a vector of weight $\alpha$. 
Define $\mathfrak{b}_\alpha=\{ x\in \mathfrak{b}\mid x_\alpha=0 \}$,
and $L_\alpha$ to be L\'evy subgroup whose Lie algebra is $\mathfrak{b}\oplus \C e_{-\alpha}$.
Starting from an orbital variety $\pi$, we distinguish two cases:
\begin{itemize}
\item $\pi\subset \mathfrak{b}_\alpha$. Then set $s_\alpha \pi=\pi$.
\item $\pi \not\subset \mathfrak{b}_\alpha$. Then let $L_\alpha$ acts by conjugation:
the top-dimensional components of $L_\alpha\cdot (\pi\cap \mathfrak{b}_\alpha)$ 
are again
orbital varieties; set $s_\alpha \pi = - \pi - \sum_{\pi'} \mu_{\alpha}{}_\pi^{\pi'} \pi'$ where
$\mu_{\alpha}{}^{\pi'}_\pi$ is the multiplicity of $\pi'$ in 
$L_\alpha\cdot (\pi\cap \mathfrak{b}_\alpha)$.
\end{itemize}
These elementary operations have a counterpart when acting on multidegrees, and a simple
calculation shows that both cases are covered by a single formula:
\newcommand{\der}{\partial}
\begin{equation}\label{funda}
s_\alpha \Psi_\pi = (-\tau_\alpha + A \der_\alpha) \Psi_\pi
\end{equation}
where $\tau_\alpha$ is the reflection orthogonal to the root $\alpha$ in $\C[\alpha_1,\ldots,\alpha_r,A]$, 
and $\der_\alpha={1\over\alpha}(\tau_\alpha-1)$
is the associated {\it divided difference operator},
whereas on the left hand side $s_\alpha$ implements right action on the $\Psi_\pi$, namely
$s_\alpha\Psi_\pi:=-\Psi_\pi-\sum_{\pi'} \mu_{\alpha}{}^{\pi'}_\pi \Psi_{\pi'}$.
One can check that $s_\alpha\mapsto -\tau_\alpha+A \der_\alpha$ is a representation of the Weyl
group $W$ on polynomials.
Note that at $A=0$, we recover the natural action of $W$ (up to a sign, with our
conventions).

\subsection{Yang--Baxter equation and integrable models}
Let us define the operator 
\begin{equation}
R_\alpha(u):={A-u s_\alpha\over A+u}
\end{equation}
which acts in the space
$V\otimes \C[\alpha_1,\ldots,\alpha_r,A]$, $u$
being a formal parameter.
Rewriting slightly the relation \eqref{funda} above we find that $\tau_\alpha$ acts as 
$R_\alpha(\alpha)$.
Using the fact that $\tau_\alpha$, just like the $s_\alpha$, satisfy the Weyl group relations,
we find that the operators $\tau_\alpha R_\alpha(\alpha)$ also satisfy those.
In the case of non-exceptional Lie algebras, there are only 2 types of edges in the Dynkin
diagram, and therefore we have Coxeter
relations of the form $(s_\alpha s_\beta)^{m_{\alpha\beta}}=1$, where 
$m_{\alpha\beta}=1,2,3,4$ depending on whether $\alpha=\beta$, 
there is no edge, a single or a double edge between
$\alpha$ and $\beta$. Writing these relations for $\tau_\alpha R_\alpha$ and eliminating the
$\tau_\alpha$, we find that relations with $m_{\alpha\beta}=1,3,4$
correspond respectively to the {\it unitarity} equation:
\begin{equation}
R_\alpha(\alpha)R_\alpha(-\alpha)=1\ ,
\end{equation}
the {\it Yang--Baxter} equation:
\begin{equation}
R_\alpha(\alpha)R_\beta(\alpha+\beta)R_\alpha(\beta)=
R_\beta(\beta)R_\alpha(\alpha+\beta)R_\beta(\alpha)
\qquad
\alpha\ \epsfig{file=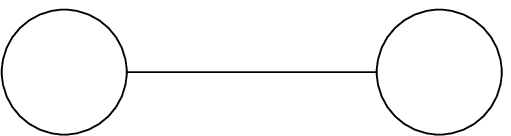, width=1cm}\ \beta
\end{equation}
and the {\it boundary Yang--Baxter} (or reflection) equation:
\begin{equation}
R_\alpha(\alpha)R_\beta(\beta+\alpha)R_\alpha(\alpha+2\beta)R_\beta(\beta)
=
R_\beta(\beta)R_\alpha(\alpha+2\beta)R_\beta(\beta+\alpha)R_\alpha(\alpha)
\qquad
\alpha\ \epsfig{file=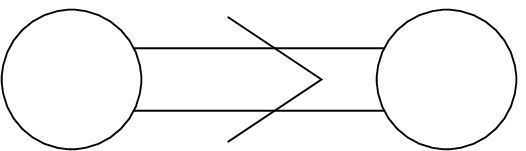, width=1cm}\ \beta
\end{equation}
whereas the case $m_{\alpha\beta}=2$ expresses a simple commutation relation for distant vertices.
Indeed one recognizes in $R_\alpha(u)$
a standard form of the rational solution of the Yang--Baxter equation, 
the parameter $u$
playing the role of difference of spectral parameters. Thus the multidegrees
$\Psi_\alpha$ are closely connected to integrable models with rational dependence on
spectral parameters, as will be discussed now.

Before doing so, let us remark
that in the special case investigated here of orbital varieties associated to $M^2=0$,
the $s_\alpha$ obey more than just the Coxeter relations. 
In the $A_r$ case they actually generate
a quotient of the symmetric group algebra $S_{r+1}$ known as
the Temperley--Lieb algebra $TL_{r+1}(2)$ (here 2 is the value of the parameter in the definition of the algebra,
as will be explained below). The same type of phenomena will be described 
for other simple Lie algebras, and will lead to variants of the Temperley--Lieb algebra; 
in particular, the ``bulk'' (i.e. everything but a finite number
of edges at the boundary) of the Dynkin diagrams being sequences of simple edges, these variants
will only differ at the level of ``boundary conditions'' of the model.

\subsection{Affinization and rational $q$KZ equation}
Let us now discuss the meaning of the equation
\begin{equation}
R_\alpha(\alpha) \Psi = \tau_\alpha \Psi
\end{equation}
where $\tau_\alpha$ is the reflection associated to the root $\alpha$ acting on the
``spectral parameters'' $\alpha_1$, $\ldots$, $\alpha_r$, $R_\alpha(\alpha)$ is a certain
linear operator defined above acting in the space $V\otimes \C[\alpha_1,\ldots,\alpha_r,A]$ and
$\Psi=\sum_\pi \pi\otimes \Psi_\pi$ is a vector in that space.

When $R_\alpha(u)$ is the $R$-matrix (or boundary $R$-matrix) of some integrable model,
such equations are satisfied by eigenvectors of the corresponding integrable transfer matrix.
More generally, these equations appear in the context of the quantum Knizhnik--Zamolodchikov
($q$KZ) equation, in connection with the representation theory of affine
quantum groups \cite{FR}. In either case, it is known that we need an additional equation 
to fix the $\Psi_\pi$ entirely.

Define $\hat{W}$ 
to be the semi-direct product of $W$ and of the weight lattice of $\mathfrak{g}$. It contains as a finite index subgroup
the usual affine Weyl group defined as the Coxeter group of the affinized Dynkin diagram.
Just like the affine Weyl group, it has a natural action on $\mathfrak{t}$ and therefore
on $\C[\alpha_1,\ldots,\alpha_r,A]$ which extends the action of $W$ generated by the reflections $\tau_i$; by definition,
in this representation, an element of the weight lattice
acts as translation in $\mathfrak{t}$ of the weight multiplied by $3A$
($3=l+\check h$ where $l=1$ is the level of the $q$KZ equation and $\check h=2$ is the dual Coxeter number of
$\mathfrak{sl}(2)$).

Then we claim that one can extend the representation of $W$ on
$V\otimes \C[\alpha_1,\ldots,\alpha_r,A]$ (the operators $\tau_\alpha R_\alpha(\alpha)$)
into a representation of $\hat{W}$, in such a way that each element of $\hat W$ is the product
of its natural action on $\C[\alpha_1,\ldots,\alpha_r,A]$
and of a $\C[\alpha_1,\ldots,\alpha_r,A]$-linear operator.
Describing here the geometric procedure that leads to this action is beyond the scope
of this paper.
The action will however be described explicitly in each of the cases below.
An important property is that
if one sets $A=0$ the representation of $\hat{W}$ factors
through the projection $\hat W \to W$. So the $\C^\times$ action actually produces
the affinization.

Imposing that $\Psi$ be invariant under the action of the whole group $\hat W$ leads to a 
full set of equations, which are precisely equivalent
to the so-called rational $q$KZ equation (or more precisely, a generalization
of it for arbitrary Dynkin diagram, the original $q$KZ equation corresponding to the case $A_r$) at
level $1$; and it turns out that they have
a unique polynomial solution of the prescribed degree (up to multiplication by a scalar).

\subsection{$q$-deformation and Razumov--Stroganov point}
The integrability suggests how to $q$-deform the above construction.
Indeed, we have considered thus far $R$-matrices that form so-called {\it rational}\/ solutions 
of the Yang--Baxter Equation,  and $\Psi$'s that are solutions of the
rational $q$KZ equation. It is known however that the trigonometric $R$-matrices are a special 
degeneration of a one-parameter family of {\it trigonometric}\/ solutions of the Yang--Baxter
Equation, depending on a parameter $q$.
Setting $q=-e^{-\hbar A/2}$, one customarily uses
exponentiated ``multiplicative" spectral parameters of the form $e^{-\hbar \alpha_i}$. 
We then look for polynomial solutions $\Psi$ of these parameters,  
to the corresponding trigonometric $q$KZ equations.
The rational solutions are then recovered from the trigonometric ones via the limit $\hbar\to0$,
at the first non-trivial order in $\hbar$. The details of the bulk and boundary $R$-matrices
will be given below for the cases $A_r$, $B_r$, $C_r$ and $D_r$. We thus obtain, for any $q$, a
representation of the group $\hat W$, the $W$ relations satisfied by the
$\tau_\alpha R_\alpha(\alpha)$ and more generally the $\hat W$ relations being undeformed.





In terms of the new variables $e^{-\hbar \alpha_i}$ living in $T$, the natural action of an element of the weight
lattice $\omega$ (as the abelian subgroup of $\hat W$) is the multiplication by $q^{6\omega}$. Since for all
simple Lie algebras, $\omega$ has half-integer coordinates, we reach the
important conclusion that when $q^3=1$, this action becomes trivial.
Therefore, all operators associated to the weight lattice by the procedure outlined in the previous
section become $\C[\alpha_1,\ldots,\alpha_r,A]$-linear
(i.e.\ correspond to finite-dimensional operators on $V$
after evaluation of the parameters $\alpha_1$, $\ldots$, $\alpha_r$, $A$).
In this case they
are simply the scattering matrices of \cite{Pasb}, and they commute with the usual (inhomogeneous)
integrable transfer matrix of the model. This implies that $\Psi$ is an eigenvector of the latter;
in fact, we can call it ``ground state eigenvector'' because in the physical situation
where the transfer matrix elements are positive, the Perron--Frobenius theorem applies and the
eigenvalue $1$ of $\Psi$ is the largest eigenvalue in modulus.

The value $q=e^{2i\pi/3}$ (also called ``Razumov--Stroganov point") is henceforth quite special and deserves 
a particular study. In particular, in the homogeneous limit where
the spectral parameters $\alpha_i$ are specialized to zero, $\Psi$ can be normalized so that its
entries are all {\it non-negative integers}, and we are interested in their combinatorial significance, 
in relation to the counting of Alternating Sign Matrices and/or Plane Partitions. 
We do not claim to have a full
understanding of the general correspondence principle between simple Lie algebras 
and these combinatorial problems, but we will perform a case-by-case study for 
$A_r$, $B_r$, $C_r$ and $D_r$. 

A last remark is in order. As we shall see, it is simple to see that
the solutions $\Psi$ to the $A$, $B$, $C$, $D$ $q$KZ equations
obey recursion relations, that allow to obtain the rank $r$ case from rank $r+1$,
hence we will content ourselves with the detailed description for $r$ with a
given parity, namely $A_{2n-1}$, $B_{2n}$, $C_{2n+1}$, $D_{2n+1}$.
  
\section{$A_r$ case}
We review the $A_r$ case, already explored in \cite{DFZJc}.
We set $\alpha_i=z_i-z_{i+1}$, $i=1,\ldots,r$. The fact that there are $r+1\equiv N$ of these new
variables $z_i$, the spectral parameters, as opposed to the $r$ simple roots,
is a reflection of the usual embedding $\mathfrak{sl}(N)\subset \mathfrak{gl}(N)$.
$\mathfrak{b}$ (resp.\ $\mathfrak{n}$) is simply the space of upper triangular
(resp.\ strictly upper triangular) matrices of size $N$, and the orbital varieties
under consideration are the irreducible components of the scheme $\{ M\in \mathfrak{n}
\mid M^2=0\}$. We also restrict ourselves to the case of $N=2n$ even, which is technically simpler.

\subsection{Orbital varieties and Temperley--Lieb algebra}
\begin{figure}
\epsfig{file=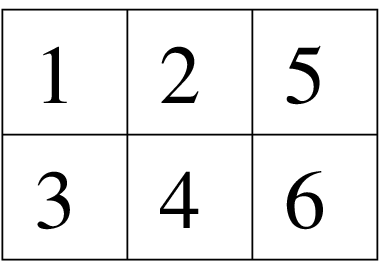,width=2cm}
\qquad\raise0.5cm\hbox{$\to$}\qquad
\epsfig{file=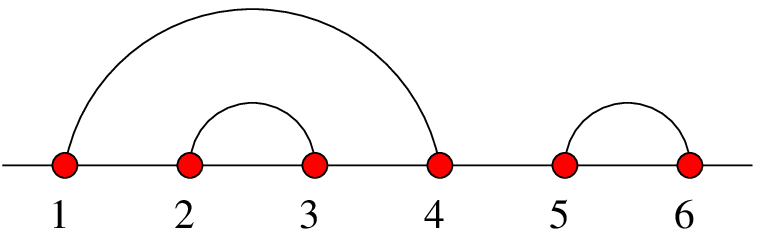,width=5cm}
\caption{A Standard Young Tableau and the corresponding link pattern.\label{linkpatt}}
\end{figure}

In general, $\mathfrak{sl}(N)$ nilpotent orbits are classified by their Jordan decomposition
type, which can be expressed as a Young diagram; the orbital varieties are then indexed by
Standard Young Tableaux (SYT). 
The condition $M^2=0$ ensures that only Young diagrams with at most
2 rows can appear (blocks in the Jordan decomposition are of size at most 2), and
it is easy to check that all orbits are in the closure of the largest orbit, whose
Young diagram is of the form $(n,n)$. 
It is convenient to describe the corresponding SYT by ``link patterns'',
that is $N$ points on a line connected in the upper-half plane
via $n$ non-intersecting arches, 
see fig.~\ref{linkpatt}. The numbers in the first (resp.\ second)
row of the SYT are the labels of the openings (resp.\ closings) of the arches. 
There are ${(2n)!\over n!(n+1)!}$ such configurations.

In this language, one has a rather convenient description of orbital varieties \cite{Roth,Mel}, 
which we mention for the sake of completeness.
Indeed, to each orbital variety $\pi$ we associate the upper triangular matrix $\pi^<$
with $\pi^<_{ij}=1$ if points labelled $i$ and $j$ are connected by an arch, $i<j$, $0$
otherwise. Then $\pi=\overline{B\cdot\pi^<}$, $B$ acting by conjugation. Equivalently, $\pi$ is
given by the following set of equations: (i) $M^2=0$ and (ii) $r_{ij}(M)\le r_{ij}(\pi^<)$,
$i,j=1,\ldots,N$, where $r_{ij}$ is the rank of the $i\times j$ lower-left rectangle.

\begin{figure}
$$e_4\ 
\raise-2.6mm\hbox{\epsfig{file=arch3.eps,width=4cm}}
=
\raise-9.2mm\hbox{\epsfig{file=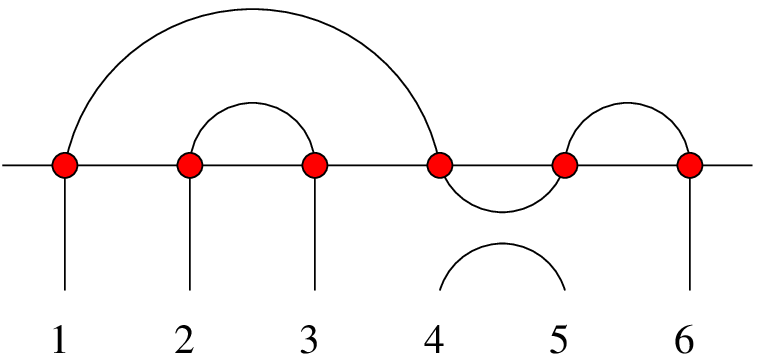,width=4cm}}
=
\raise-2.6mm\hbox{\epsfig{file=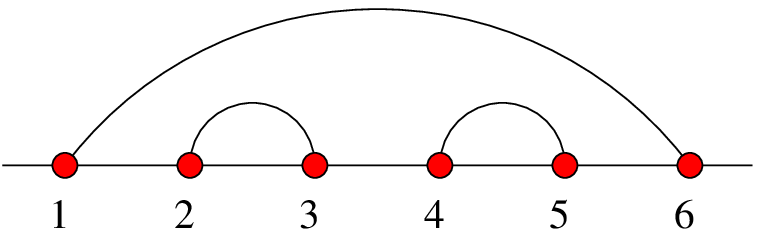,width=4cm}}
$$
$$
e_2\ 
\raise-2.6mm\hbox{\epsfig{file=arch3.eps,width=4cm}}
=
\raise-9.2mm\hbox{\epsfig{file=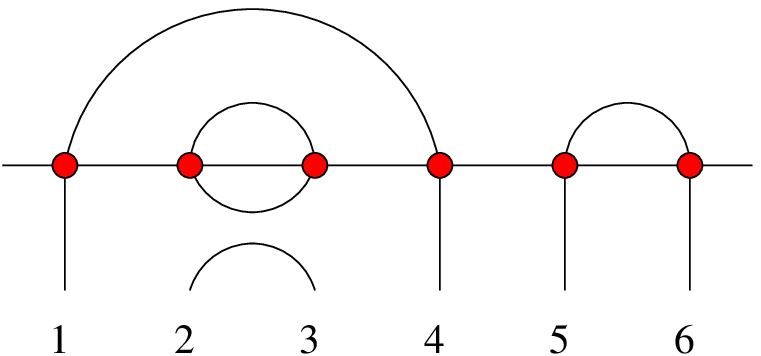,width=4cm}}
=
\beta\ 
\raise-2.6mm\hbox{\epsfig{file=arch3.eps,width=4cm}}
$$
\caption{Action of the Temperley--Lieb algebra $TL(\beta)$ on link patterns.\label{tla}}
\end{figure}

It is equally simple to describe the action of the Weyl group, namely the symmetric
group $S_N$. Rather than the generators corresponding to the simple roots:
$s_i\equiv s_{\alpha_i}$, $i=1,\ldots,r$ used so far, it proves simpler to consider
the action of the projectors $e_i=1-s_i$ in the symmetric group algebra.
The operator $e_i$ acts on link patterns $\pi$ by connecting the arches ending at $i$ and $i+1$ 
and creates a new little arch between these 2 points;
this action is described on Fig.~\ref{tla}. When a closed loop is formed,
it is erased but contributes a weight $\beta=2$.  
The $q$-deformed version of this is obtained by attaching a weight $\beta=-(q+q^{-1})$ to
each erased loop, thus leading to the following (pictorially clear) relations:
\begin{equation}\label{tlarel}
e_i^2=\beta e_i\qquad e_i=e_i e_{i\pm 1}e_i\qquad [e_i,e_j]=0\quad |i-j|>1
\end{equation}
all indices taking values in $1,\ldots,r$. These are the defining relations
of the Temperley--Lieb algebra $TL_{r+1}(\beta)$. When $q=-1$, i.e. $\beta=2$, it is simply 
a quotient of the symmetric group algebra. Alternatively, the deformed generators
$s_i=-q^{-1}-e_i$ satisfy the usual relations of the Hecke algebra (of which the Temperley--Lieb
algebra is a quotient).

In what follows, one special element of $TL_N(\beta)$ will be needed: 
it is the cyclic rotation $S$. 
Its effect is to rotate the endpoints of the link patterns: $1\to2\to\cdots\to N\to1$ 
without changing their connectivity.
It can also be expressed as: $S=q^{n-2} s_1 \cdots s_{N-1}$.

\subsection{$q$KZ equation}
For each simple root $\alpha_i$, we have the trigonometric $R$-matrix:
\begin{equation}\label{qRmat}
R_i(w)\equiv R_{\alpha_i}(w)={(qw-q^{-1})+(w-1)e_i\over q-q^{-1}w}, 
\end{equation}
where the $e_i=-q^{-1}-s_i$ generate $TL_N(\beta)$ and
act in the space of link patterns
as explained above. We first write the system of equations:
\begin{equation}\label{qkza}
R_i(w_{i+1}/w_{i})\Psi=\tau_i \Psi \qquad i=1,\ldots,N-1
\end{equation}
where $\tau_i\equiv \tau_{\alpha_i}$ acts by interchanging multiplicative spectral parameters $w_i:=e^{-\hbar z_i}$
and $w_{i+1}$
in the polynomial $\Psi$ of the $w$'s, homogeneous of degree $n(n-1)$.

These equations are supplemented by the ``affinized'' equation satisfied by $\Psi$.
Since the affine Dynkin diagram $A_r^{(1)}$ is a circular chain,
this equation quite naturally involves the cyclic rotation $S$. 
Define the operator $\rho$ on $\C[w_1,\ldots,w_N]$
which shifts the variables
$w_i$ according to the rule: $w_i\to w_{i+1}$, $i=1,\ldots,N-1$ and $w_N\to q^6 w_1$.
Then the additional equation is
\begin{equation}\label{qkzaa}
q^{3(n-1)}S^{-1} \Psi=\rho\Psi
\end{equation}
Together with this equation, the above system forms the so-called level one $q$KZ equation.

We claim that the ${\bf R}_i:=\tau_i R_i(w_{i+1}/w_i)$ and ${\bf S}:=q^{3(1-n)}\rho S$ 
generate together $\hat W$. In order to see
that, it is sufficient to build the $N$ generators ${\bf T}_i$ of the abelian subgroup (the lattice of weights).
They are given by
${\bf T}_i={\bf R}_{i-1}{\bf R}_{i-2}\cdots {\bf R}_1 {\bf S} {\bf R}_{N-1}\cdots {\bf R}_{i+1}
{\bf R}_i$,
$i=1,\ldots,N$. 
The original definition of the $q$KZ equation is in fact the eigenvector equation for these 
``scattering'' matrices;
with reasonable assumptions it is equivalent to the above system.
Also, note that if one defines ${\bf R}_N:={\bf S}^{-1} {\bf R}_1 {\bf S}$, 
then the ${\bf R}_i$, $i=1,\ldots,N$ generate
the usual affine Weyl group (a subgroup of order $N$ of $\hat W$).

The minimal degree polynomial solution of the level one $q$KZ equation was obtained in
\cite{Pas,DFZJc}, and is characterized by its ``base'' entry $\Psi_{\pi_0}$ corresponding
to the link pattern $\pi_0$ that connects points $i\leftrightarrow 2n+1-i$, with the value
\begin{equation}
\Psi_{\pi_0}=\prod_{1\leq i<j\leq n} (qw_i-q^{-1}w_j)\prod_{n+1\leq i<j\leq 2n}(qw_i-q^{-1}w_j)
\end{equation}
in which all factors are a direct consequence of the $\tau_i\Psi=R_i\Psi$ equations. 
It is then easy to prove that all the other entries of $\Psi$ may be obtained from $\Psi_{\pi_0}$
in a triangular way.

{\it Example:} at $N=6$, there are 5 link patterns. The minimal degree polynomial
solution of the level one $q$KZ equation reads:
\begin{eqnarray*}
&&\Psi_{\epsfig{file=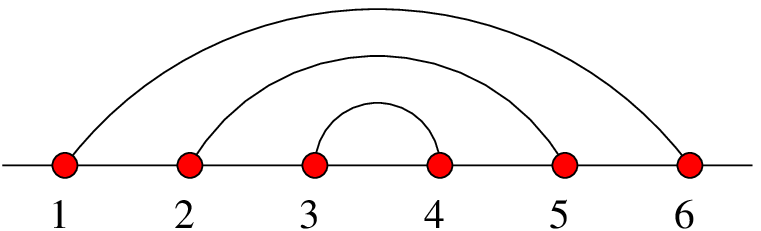,width=1.8cm}}
=(qw_1-q^{-1}w_2)(qw_2-q^{-1}w_3)(qw_1-q^{-1}w_3)
(qw_4-q^{-1}w_5)(qw_5-q^{-1}w_6)(qw_4-q^{-1}w_6)\\
&&\Psi_{\epsfig{file=arch1.eps,width=1.8cm}}
=(qw_1-q^{-1}w_2)(qw_3-q^{-1}w_4)(qw_5-q^{-1}w_6)\\
&&\times
\Big((w_1+w_2)(q^2w_3w_4-q^{-2}w_5w_6)-(w_3+w_4)(q^4w_1w_2-q^{-4}w_5w_6)
+(w_5+w_6)(q^2w_1w_2-q^{-2}w_3w_4)\Big)\\
&&\Psi_{\epsfig{file=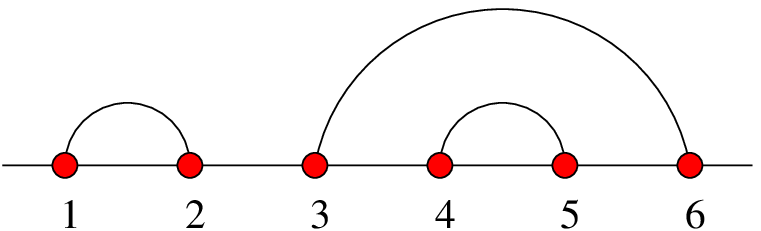,width=1.8cm}}
=(qw_2-q^{-1}w_3)(qw_2-q^{-1}w_4)(qw_3-q^{-1}w_4)
(qw_5-q^{-1}w_6)(q^{-2}w_6-q^{2}w_1)(q^{-2}w_5-q^{2}w_1)\\
&&\Psi_{\epsfig{file=arch3.eps,width=1.8cm}}
=(qw_1-q^{-1}w_2)(qw_3-q^{-1}w_4)(qw_4-q^{-1}w_5)
(qw_3-q^{-1}w_5)(q^{-2}w_6-q^{2}w_1)(q^{-2}w_6-q^{2}w_2)\\
&&\Psi_{\epsfig{file=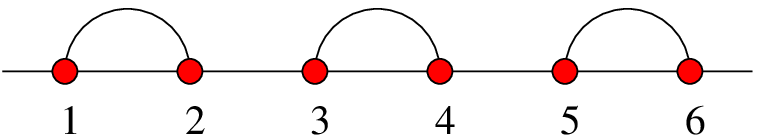,width=1.8cm}}
=(qw_2-q^{-1}w_3)(qw_4-q^{-1}w_5)(q^{-2}w_6-q^{2}w_1)\\
&&\times
\Big((q^3w_1+q^{-3}w_6)(q^2w_2w_3-q^{-2}w_4w_5)-(w_2+w_3)(qw_1w_6-q^{-1}w_4w_5)
-(w_4+w_5)(qw_2w_3-q^{-1}w_1w_6)\Big)\\
\end{eqnarray*}
Performing the rational limit $\hbar\to 0$, $z_i=e^{-\hbar w_i}$, $q=-e^{-\hbar A/2}$ yields the following multidegrees:
\begin{eqnarray*}
&&\Psi_{\epsfig{file=arch0.eps,width=1.8cm}}
=(A+z_1-z_2)(A+z_2-z_3)(A+z_1-z_3)
(A+z_4-z_5)(A+z_5-z_6)(A+z_4-z_6)\\
&&\Psi_{\epsfig{file=arch1.eps,width=1.8cm}}
=(A+z_1-z_2)(A+z_3-z_4)(A+z_5-z_6)\Big(4A^3 + 3A^2(z_1+z_2-z_5-z_6) +\\
&&+ A(2(z_1z_2-2z_3z_4-z_1z_5-z_2z_5-z_1z_6-z_2z_6+z_5z_6)+ (z_3+z_4)(z_1+z_2+z_5+z_6))\\
&&+ (z_1+z_2)(z_5z_6-z_3z_4)+(z_3+z_4)(z_1z_2-z_5z_6)+(z_5+z_6)(z_3z_4-z_1z_2)\Big)\\
&&\Psi_{\epsfig{file=arch2.eps,width=1.8cm}}
=(A+z_2-z_3)(A+z_2-z_4)(A+z_3-z_4)
(A+z_5-z_6)(2A+z_1-z_6)(2A+z_1-z_5)\\
&&\Psi_{\epsfig{file=arch3.eps,width=1.8cm}}
=(A+z_1-z_2)(A+z_3-z_4)(A+z_4-z_5)
(A+z_3-z_5)(2A+z_1-z_6)(2A+z_2-z_6)\\
&&\Psi_{\epsfig{file=arch4.eps,width=1.8cm}}
=(A+z_2-z_3)(A+z_4-z_5)(2A+z_1-z_6)\Big(5A^3 + 3A^2(z_1+z_2+z_3-z_4-z_5-z_6) +\\
&&+A(2z_1(z_2+z_3-z_6)+z_2z_3+z_4z_5-(z_2+z_3)z_6+(z_4+z_5)(2z_6-z_1-z_2-z_3))\\
&&+(z_1+z_6)(z_2z_3-z_4z_5)+(z_2+z_3)(z_4z_5-z_1z_6)+(z_4+z_5)(z_1z_6-z_2z_3)\Big)\\
\end{eqnarray*}
and in particular the degrees $1,4,4,4,10$ respectively, upon taking $z_i=0$ and $A=1$.

\subsection{Razumov--Stroganov point and ASM}
At $q=e^{2i\pi/3}$, $\Psi$ becomes the ground state eigenvector of the integrable transfer matrix
with periodic boundary conditions and inhomogeneities $w_1$, $\ldots$, $w_N$,
or equivalently of the scattering matrices
${\bf T}_i=R_{i-1}(w_{i-1}/w_i)\cdots R_1(w_1/w_i) S R_{N-1}(w_{N-1}/w_i)\cdots
R_i(w_{i+1}/w_i)$.
Consider now the particular case
$w_1=\cdots=w_N=1$, when $\Psi$ is the Perron--Frobenius eigenvector
of the Hamiltonian $H=e_1+\cdots+e_N$ where $e_N=S^{-1} e_1 S$. 
Note that the periodic boundary conditions mean that
$H$ is cyclic-invariant: $SH=HS$.
Normalizing $\Psi$ so that
its smallest entry $\Psi_{\pi_0}$ is $1$, we have the following
\begin{theorem*}
\cite{DFZJ}
The sum of entries $\sum_\pi \Psi_\pi$ is equal to the number of Alternating Sign
Matrices, 
$A(n)$. 
\end{theorem*}
The result of \cite{DFZJ} is actually much more general, as the sum $\sum_\pi \Psi_\pi$
was evaluated in the presence of all the spectral parameters $w_i$, and identified
with proper normalization to the so-called Izergin--Korepin determinant \cite{Iz,Kor},
also equal to a particular Schur function \cite{Oka}.
Still unproven, however, is the
\begin{conjecture*}
\cite{BdGN}
The largest entry of $\Psi$, with arches connecting consecutive points, is $A(n-1)$.
\end{conjecture*}

For instance, plugging $w_i=1$ and $q=e^{2i\pi/3}$
into the above example, we get for $N=6$, $\Psi=(1,2,1,1,2)$ and $\sum_\pi \Psi_\pi=7=A(3)$, the
total number of $3\times 3$ ASMs.

\section{$B_r$ case}
We now develop the $B_r$ case, which allows us to recover and interpret
geometrically the results of \cite{DF}. We concentrate on the even case $r=2n$.
We parametrize as usual the roots 
$\alpha_i=z_i-z_{i+1}$ for $i=1,2,\ldots,r-1$ and $\alpha_r=z_r$. 

We consider matrices that square to zero
in the fundamental representation of dimension $N=2r+1$: a possible choice
is to select upper triangular matrices satisfying $M^TJ+JM=0$, $J$ antidiagonal matrix with 1's on the
second diagonal. 
It turns out that the orbital varieties are indexed by the same link patterns as before, of size $r$; and that the
Weyl group representation is actually a representation of the same quotient, the Temperley--Lieb algebra
$TL_r(\beta)$, the additional reflection $s_r$ being represented by a multiple of the identity.

\subsection{$B$-type $q$KZ equation}
According to the dicusssion above, the B $q$KZ system reads:
\begin{eqnarray}
R_i(w_{i+1}/w_i)\Psi&=&\tau_i \Psi, \quad i=1,2,...,r-1\label{firBx} \\ 
w_r^{-m_r}{q^{-1}w_r-q\over q^{-1}-q w_r} \Psi&=&\tau_r \Psi\label{firB}
\end{eqnarray}
where $\tau_r$ stands for the inversion of the last spectral parameter, namely
$\tau_r\Psi(w_1,...w_{r-1},w_r)=  \Psi(w_1,...,w_{r-1},1/w_r)$
and $m_r$ is the degree of $\Psi$ in $w_r$.

Finally, these equations are to be supplemented by the affinization relation. 
The latter is expressed by considering the reflection with respect to the extra root
$z_1$. One finds that
\begin{equation}\label{secB}
(q^3w_1)^{-m_1} {q^{-2}-q^2 w_1\over q w_1-q^{-1}}
\Psi(w_1,w_2,...,w_r)=
\Psi\Big({1\over q^6w_1},w_2,...,w_{r}\Big)
\end{equation}
where $m_1$ is the degree of $\Psi$ in $z_1$.

Introducing the boundary operators ${\bf K}_1$ and ${\bf K}_r$ so that Eqs.~(\ref{firB}--\ref{secB}) reduce to 
${\bf K}_1 \Psi={\bf K}_2 \Psi=\Psi$, as well as the usual ${\bf R}_i=\tau_i R_i(w_{i+1}/w_i)$,
the generators of the weight lattice (as abelian subgroup of $\hat W$)
are: (i) ${\bf T}_i={\bf R}_i{\bf R}_{i+1}\cdots {\bf R}_{r-1}{\bf K}_r
{\bf R}_{r-1}\cdots$ ${\bf R}_1 {\bf K}_1 {\bf R}_1\cdots{\bf R}_{i-1}$
that implements $w_i\to q^6 w_i$ and (ii) one additional generator implementing $w_i\to q^3 w_i$ simultaneously
for all $i$. The latter is a combination of {\bf R} and {\bf K} matrices as well as an additional
operator implementing the reflection $w_i\leftrightarrow q^{-3}/w_{r+1-i}$ for all $i$.

The minimal polynomial solution to the system (\ref{firBx}--\ref{secB})
has degree $m_1=m_r=r-1=2n-1$ in each spectral parameter
and total degree $n(3n-1)$. As before it has a simple factorized base entry
\begin{equation}
\Psi_{\pi_0}=C\prod_{1\leq i<j\leq n}(qw_i-q^{-1}w_j)(q^{-2}-q^{2}w_iw_j)
\prod_{n+1\leq i<j\leq 2n} (qw_i-q^{-1}w_j)(qw_iw_j-q^{-1})
\end{equation}
where $C=2^n\prod_{i=1}^r (q w_i-q^{-1})$ is a common (symmetric) factor to all entries of $\Psi$. 
All other entries may be obtained from this one in a triangular manner.

{\it Example:} For $B_4$, there are 2 link patterns as for the case $A_3$.
The minimal degree polynomial
solution of the level one $B_4$ $q$KZ equation reads:
\begin{eqnarray*}
&&\Psi_{\epsfig{file=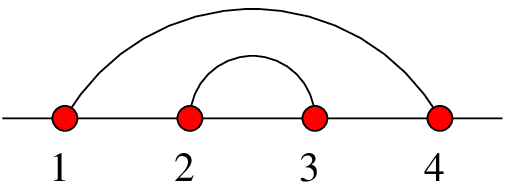,width=1.2cm}}
=C(qw_1-q^{-1}w_2)(q^{-2}-q^{2}w_1w_2)(qw_3-q^{-1}w_4)(qw_3w_4-q^{-1})\\
&&\Psi_{\epsfig{file=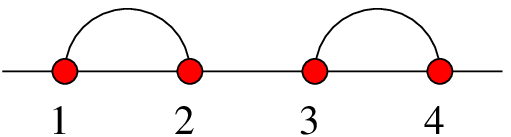,width=1.2cm}}
=C(qw_2-q^{-1}w_3)(q^{-1}w_1-qw_1w_2w_3-q^{-5}w_4-qw_1^2w_4+(q^{-1}-q)w_1(w_2+w_3)w_4\\
&&\ \ \ +q^{-1}w_2w_3w_4+q^5 w_1^2 w_2w_3w_4+q^{-1}w_1w_4^2-qw_1w_2w_3w_4^2)\\
\end{eqnarray*}
As before, we get the corresponding multidegrees upon taking the rational limit,
with the result:
\begin{eqnarray*}
&&\Psi_{\epsfig{file=arch0b.eps,width=1.2cm}}
=C'(A+z_1-z_2)(2 A + z_1+ z_2) (A + z_3 - z_4) (A + z_3 + z_4)\\
&&\Psi_{\epsfig{file=arch1b.eps,width=1.2cm}}
=C'(A+z_2-z_3)\Big(5A^3+3A^2 (2z_1 + z_2 + z_3)+A(2z_1^2 + 3z_1(z_2 +z_3) +z_2z_3 -z_4^2)\\
&&+ (z_2+z_3)(z_1^2-z_4^2)\Big)\\
\end{eqnarray*}
with $C'=4(A+z_1)(A+z_2)(A+z_3)(A+z_4)$;
hence the degrees $4\times2,4\times5$ for $A=1$ and $z_i=0$.



\subsection{RS point, VSASM and CSTCPP}
As explained in Sect.~2, the case $q=e^{2i\pi/3}$ is special in that the problem admits a
transfer matrix, and its solution $\Psi$ in the homogeneous
limit where all $w_i=1$ is the groundstate of a Hamiltonian
\begin{equation}
H_B=e_1+e_2+...+e_{N-1}
\end{equation}
which is the open boundary version of the $A_r$ Hamiltonian $H$.

As shown in \cite{DFb}, at the RS point $q=e^{2i\pi/3}$, and in the homogeneous limit
where $w_i=1$ for all $i$, and in which $\Psi$ is normalized so that its smallest entry is 
$\Psi_{\pi_0}=1$, we have the following
\begin{theorem*}
\cite{DF}
The sum of entries $\sum_\pi \Psi_\pi$ is equal to the number of Vertically Symmetric
Alternating Sign Matrices (VSASM), $A_V(2n+1)$. 
\end{theorem*}
This was actually proved in the same spirit as for the $A_r$ case, by identifying the 
sum of components including all spectral parameters $w_i$ as yet another determinant,
which takes the form of a particular symplectic Schur function.
A similar result holds for the case of odd $r=2n-1$, namely once properly normalized,
the sum of entries $\sum_\pi \Psi_\pi$ is equal to an integer we call $A_V(2n)$ by analogy.
It turns out that $A_V(2n)=N_8(2n)$ is the number of Cyclically Symmetric Transpose Complement
Plane Partitions (CSTCPP) in an hexagon of size $2n\times 2n\times 2n$ \cite{Bre}.
The numbers $A_V(i)$ both have determinant
formulae, namely
$A_V(2n)=\det{i+j\choose 2i-j}_{0\leq i,j\leq n-1}$,
and
$A_V(2n+1)=\det{i+j+1\choose 2i-j}_{0\leq i,j\leq n-1}$.

As in the $A$ case, we have the
\begin{conjecture*}\cite{BdGN}
The largest entry of $\Psi$, with arches connecting consecutive points, is $A_V(r)$.
\end{conjecture*}

Example: for $r=2n=4$, taking $w_i\to 1$ and $q=e^{2i\pi/3}$ in the above expressions, we get
the components
$\Psi=(1,2)$, which sum to $3=A_V(5)$, the number of $5\times 5$ VSASMs, 
and the maximal entry of $\Psi$ is $2=N_8(4)$.

\section{$C_r$ case}
The simple roots of $C_r$ are
$\alpha_i=z_i-z_{i+1}$, $i=1,2,\ldots,r-1$ and $\alpha_r=2z_r$. 
We concentrate on the odd case $r=2n+1$, and consider
the fundamental representation of dimension $N=2r$. One choice is to select
upper triangular matrices satisfying $M^T J+JM=0$, $J$ antidiagonal matrix with $1$'s (resp.\ $-1$'s) 
in the upper (resp.\ lower) triangle.

\subsection{Orbital varieties and $C$-type Temperley--Lieb algebra}
There are $r\choose\lfloor{r+1\over2}\rfloor$
orbital varieties, which are now indexed by {\it open link patterns}, that is configurations of $r$ points on a line
connected in the upper-half plane either in pairs via (closed) arches or to infinity via half-lines (open arches).

The representation of the Weyl group on these open link patterns takes the form of a modified Temperley--Lieb algebra.
We describe now its $q$-deformed version, $CTL(\beta)$ (see also \cite{Gr} for other variants of Temperley--Lieb algebra).
The generators $e_1,e_2,\ldots,e_{r-1}$ obey the standard $TL(\beta)$ relations \eqref{tlarel}
and the additional ``boundary" generator $e_r$ satisfies:
$e_r^2=\beta e_r$, $e_{r-1}e_re_{r-1}=2 e_{r-1}$.

\begin{figure}
\epsfig{file=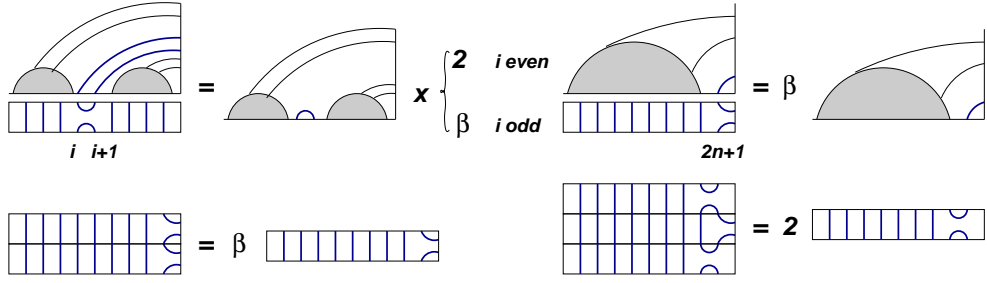,width=13cm}
\caption{The rule for erasing arches at infinity when acting with $e_i$: they are replaced by a factor 
$2$ (resp. $\beta$)
according to whether the index $i$ is even (resp. odd). We have also represented the 
case $i=2n+1$ (first line), and the resulting boundary relations $e_r^2=\beta e_r$ and $e_{r-1}e_re_{r-1}=2 e_{r-1}$
(second line).\label{crules}}
\end{figure}

These generators act on open link patterns as follows.
Open link patterns are represented with their open arches
connected to a vertical line on the right. The $e_i$, $i=1,2,...,r-1$
act as usual, and $e_r$ like the left half of an $e$, connecting the point $2n+1$
to the vertical line (first line of Fig.~\ref{crules}). The rule is that any loop
may be erased and replaced by a factor $\beta$. Moreover, whenever a connection between 
points on the vertical line (consecutive open arches) is created, they may also be erased 
and replaced by a factor
$\beta$ (resp. $2$) if this is created by the action of some $e_{2i-1}$ (resp. $e_{2i}$). 
As $r$ is odd, the loop created by $e_r^2$ yields a weight $\beta$, while that created
by $e_{r-1}e_re_{r-1}$ yields a weight $2$, hence the result $2e_{n-1}$ 
(second line of Fig.~\ref{crules}).

We shall also need an
additional operator $e_1'$ 
satisfying the relations:
$(e_1')^2=\beta e_1'$ and $e_1e_1'=e_1'e_1=e_1'e_2e_1'-e_1'=e_2e_1'e_2-e_2=0$. It is defined as $e'_1=s e_1 s$,
where $s$ is the involution acting on link patterns as follows:
(i) $s\pi=\pi$ if the arch connected to point $1$ is open,
and (ii) $s\pi=-\pi+\pi'$ otherwise, 
where $\pi'$ is the link pattern in which the closed arch connected to $1$ is cut into two open arches.

\subsection{$C$-type $q$KZ equation}

To each simple root we attach
respectively the standard trigonometric
$R$-matrices $R_i(w_{i+1}/w_i)$, $i=1,2,\ldots,r-1$ of Eq.~\eqref{qRmat}, 
and the boundary $R$-matrix $R_r(1/w_r^2)\equiv R_{\alpha_r}$, with the same expression.

The level one $C$ $q$KZ equation consists of the following system
\begin{eqnarray}
R_i(w_{i+1}/w_i)\Psi&&=\tau_i\Psi \label{CqKZa}\\
w_r^{-m_r}R_r(1/w_r^2)\Psi&&=\tau_r\Psi\label{CqKZb}
\end{eqnarray}
where as usual $\tau_i$ acts by interchanging the spectral parameters $w_i$ and $w_{i+1}$,
$i=1,2,...,r-1$ and $\tau_r$ acts on $\Psi$ by letting $w_r\to 1/w_r$,
and $m_r$ is the degree of $\Psi$ in $w_r$.

These are finally supplemented by the affinization relation, obtained by considering an extra root, say
$\alpha_1'=-z_1-z_2$, and the associated boundary operator $R_1'(q^6w_1w_2)$:
\begin{equation}
R_1'(q^6w_1w_2)\Psi=\tau_1'\Psi\label{CqKZc}
\end{equation}
where $\tau_1'$ interchanges $w_2$ and $1/(q^6 w_1)$, and $R'_1$ is of the form of Eq.~\eqref{qRmat} with
$e'_1$ in place of $e_i$. Using $R'_1(w)=s R_1(w) s$, the relation can also be recast into 
\begin{equation}
(q^3 z_1)^{-m_1}
s\Psi(w_1,\ldots,w_r)=\Psi\Big({1\over q^6 w_1},w_2,\ldots,w_r)\label{CqKZd}
\end{equation}

The generators of the weight lattice (as abelian subgroup of $\hat W$) are very similar to the generators (i)
of the case $B_r$:
the only change concerns the boundary operators ${\bf K}_1$ and ${\bf K}_r$ now implementing Eqs.~\eqref{CqKZb}
and \eqref{CqKZd}.

The polynomial solution $\Psi$ to the level one $C_r$ $q$KZ system has degree $m_1=m_r=2n$ in each variable, total degree
$n(2n+1)$ and base entry
\begin{equation}
\Psi_{\pi_0}=\prod_{1\leq i<j\leq 2n+1} (qz_i-q^{-1}z_j) 
\end{equation}
and all the other entries of $\Psi$ may be obtained in a triangular way from this one.

Example: for $r=3$, we have the following minimal polynomial solution to the level one $C_3$ $q$KZ
system:
\begin{eqnarray*}
&&\Psi_{\epsfig{file=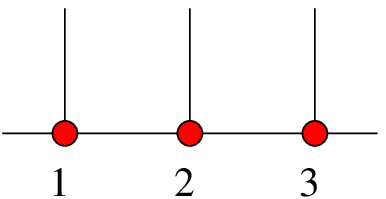,width=0.9cm}}
=(qw_1-q^{-1}w_2)(qw_1-q^{-1}w_3)(qw_2-q^{-1}w_3)\\
&&\Psi_{\epsfig{file=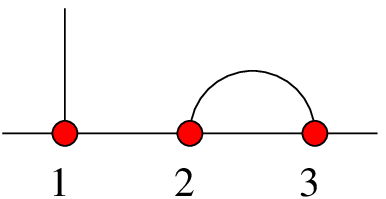,width=0.9cm}}
=(qw_1-q^{-1}w_2)(q^2 w_1w_2-q^{-2})(q^{-1}-q w_3^2)\\
&&\Psi_{\epsfig{file=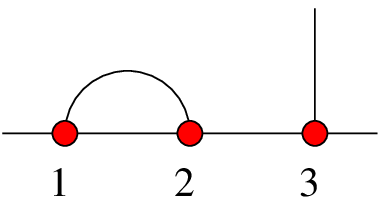,width=0.9cm}}
=(q^3w_1^2-q^{-3})(qw_2-q^{-1}w_3)(qw_2w_3-q^{-1}) \\
\end{eqnarray*}
which, upon taking the rational limit yields the multidegrees:
\begin{eqnarray*}
&&\Psi_{\epsfig{file=arch0c.eps,width=0.9cm}}
=(A+z_1-z_2)(A+z_1-z_3)(A+z_2-z_3)\\
&&\Psi_{\epsfig{file=arch1c.eps,width=0.9cm}}
=(A+z_1-z_2)(2A+z_1+z_2)(A+2z_3)\\
&&\Psi_{\epsfig{file=arch2c.eps,width=0.9cm}}
=(3A+2z_1)(A+z_2-z_3)(A+z_2+z_3) \\
\end{eqnarray*}
and the degrees $\Psi=(1,2,3)$ for $A=1$ and $z_i=0$.

\subsection{RS point and CSSCPP}
At the point $q=e^{2i\pi/3}$, $\Psi$ may be viewed as the ground state eigenvector 
of a transfer matrix, corresponding in the homogeneous limit to the
Hamiltonian 
\begin{equation}
H_C={e_1+e_1'\over 2}+\sum_{i=2}^{r-1}e_i +e_r
\end{equation}

Normalizing $\Psi$ so that its smallest entry $\Psi_{\pi_0}=1$, we have been able to compute the sum of entries to 
be $A(n)A(n+1)$.
In the case of even $r=2n$, the above may be repeated almost identically:
in the presence of spectral
parameters, the even case may be recovered from the odd one
by taking $w_{2n+1}\to -q^{-1}$, and dividing out the result by 
$\prod_{1\leq i\leq 2n} (1+q^3w_i)$. Indeed, this specialization leaves us with 
only non-vanishing components whith an open arch at the rightmost point, in bijection
with open link patterns with that point erased, hence the projection onto the case of size one less.
This leads us to the
\begin{conjecture*}
\begin{equation}
\sum_\pi \Psi_\pi =A(\lfloor r/2\rfloor)A(\lceil r/2 \rceil)
\end{equation}
\end{conjecture*}
Note that the sum in the even case, $A(n)^2$, also counts the 
Cyclically Symmetric Self-Complementary Plane Partitions (CSSCPP) in an 
hexagon of size $2n\times 2n\times 2n$ \cite{Bre}. Also note the determinant formulae
$A(n)^2=\det\Big({i+j\choose 2i-j-1}+{i+j+1\choose 2i-j}\Big)_{0\leq i,j\leq n-1}$ and
$A(n)A(n+1)=\det\Big({i+j+1\choose 2i-j}+{i+j+2\choose 2i-j}\Big)_{0\leq i,j\leq n-1}$.

Furthermore, consider the left eigenvector $v$ of $H_C$ with the same eigenvalue ($r$ for $r$ odd, $r+1/2$ for $r$ even). Normalize $v$ so that its
entries are coprime positive integers.
We have found empirically the following
\begin{conjecture*}
\begin{equation}
\sum_\pi v_\pi \Psi_\pi = A(r)\ .
\end{equation}
\end{conjecture*}

Finally, we formulate the
\begin{conjecture*}
The largest entry of $\Psi$ for $C_r$ is the sum of entries for $C_{r-1}$.
\end{conjecture*}

Example: at $r=5$, $\Psi=(1,2,3,3,0,1,4,0,0,0)$, $v=(48,36,28,34,24,23,25,18,17,14)$, 
$\sum_\pi \Psi_\pi=14=2\times 7=A(2)A(3)$, $\sum_\pi v_\pi \Psi_\pi=429=A(5)$, and the maximal entry of $\Psi$
is $4=A(2)^2$.

\section{$D_r$ case}
The simple roots of $D_r$ are 
$\alpha_i=z_i-z_{i+1}$
for $i=1,2,\ldots,n-1$ and $\alpha_r=z_{r-1}+z_r$.
We concentrate on the odd case $r=2n+1$, and consider again the fundamental 
representation of dimension $N=2r$. Just like in the $B_r$ case,
one choice is to select
upper triangular matrices satisfying $M^T J+JM=0$, $J$ antidiagonal matrix with $1$'s on the
second diagonal.

\subsection{Orbital varieties and $D$-type Temperley--Lieb algebra}
Just as in the case $C$, there are 
$r\choose\lfloor{r+1\over2}\rfloor$ orbital varieties, indexed by open link patterns.

We now deal with $D$-type Temperley--Lieb algebras, denoted $DTL(\beta)$,
with generators $e_i$, $i=1,2,...,r-1$ obeying the $TL(\beta)$ relations \eqref{tlarel} 
and an extra generator $e_{r-1}'$, satisfying the relations:
\begin{equation}
(e_{r-1}')^2=\beta e_{r-1}, \qquad 
e_{r-1}e_{r-1}'=e_{r-1}'e_{r-1}=e_{r-2}e_{r-1}'e_{r-2}-e_{r-2}=e_{r-1}'e_{r-2}e_{r-1}'-e_{r-1}'=0
\end{equation}

\begin{figure}
\epsfig{file=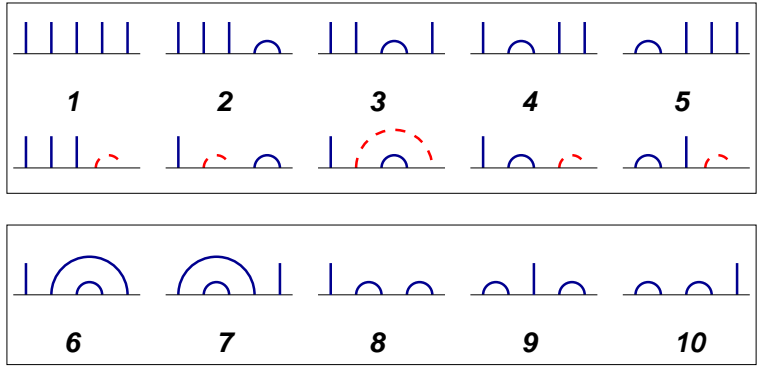,width=10cm}
\caption{The ten open link patterns for $D_5$. In the second line, we have transformed
the open link patterns by connecting the two rightmost open arches into a (dashed) arch.
The involution $s$ simply switches the color of the rightmost arch (if it is closed) in this representation,
namley exchanges $1\leftrightarrow 2$, $3\leftrightarrow 6$, $4\leftrightarrow 8$, $5\leftrightarrow 9$, 
and leaves $7$ and $10$ invariant (as their rightmost arch is open).\label{linkpaD}}
\end{figure}

These operators act on open link patterns as follows. The $e_i$, $i=1,2,\ldots,r-1$ act in the usual way,
by creating a little arch between points $i$ and $i+1$ and by gluing the two former points. To describe
the action of $e_{r-1}'$, let us first connect the open arches of the open link patterns by pairs
of consecutive open arches from the left to the right, and represent the newly formed arches in a
different color (dashed lines, cf Fig.~\ref{linkpaD} for the $D_5$ example). 
We then define an involution $s$ on open
link patterns that simply switches the color (solid $\leftrightarrow$ dashed) of the rightmost arch
if it is closed, and leaves it invariant if it is open.  Then $e_{r-1}'=s e_{r-1}s$.

Finally, we introduce an extra boundary operator $e_0$, which is the right half of an $e$ (like a reflected $e_r$ of $C_r$), 
with its open end connected to the vertical line, and acts as such, with the same rules as for $C_r$,
but upon reflection of indices $i\leftrightarrow r-i$.
It satisfies the relations: $e_0^2=\beta e_0$ and $e_1e_0e_1=2 e_1$.

\subsection{$D$-type $q$KZ equation}
We associate to the roots the $R$-matrices $R_i(w_{i+1}/w_i)$ of Eq.~\eqref{qRmat}, and $R_r(1/(w_r w_{r-1}))$
defined by the same equation in which $e_i$ is replaced with $e'_{r-1}$, so that $R_r(w)=s R_{r-1}(w) s$.

The level one $D$ $q$KZ equation consists of the following system
\begin{eqnarray*}
R_i(w_{i+1}/w_i)\Psi&&=\tau_i\Psi, \qquad i=1,2,...,r-1\\
R_r(1/(w_rw_{r-1}))\Psi&&=\tau_{r-1}'\Psi
\end{eqnarray*}
where as usual $\tau_i$ acts by interchanging the spectral parameters $w_i$ and $w_{i+1}$,
$i=1,2,...,r-1$ and $\tau_r'$ acts on $\Psi$ by interchanging $w_{r-1}$ and $1/w_r$.
Upon using the above relation $e_{r-1}'=s e_{r-1}s$, the latter equation may be equivalently
replaced by
\begin{equation}
z_r^{-m_r} s\Psi(z_1,\ldots,z_r)=\Psi\Big(z_1,\ldots,z_{r-1},{1\over z_r}\Big)
\end{equation}

These are finally supplemented by the affinization relation, obtained by considering the extra root
$\alpha_0=-2z_1$, and the associated boundary operator $R_0(q^6w_1^2)$ involving the
extra operator $e_0$:
\begin{equation}
w_1^{-m_1}R_0(q^6 w_1^2)\Psi=\tau_0\Psi
\end{equation}
where $\tau_0 f(w_1)=f( 1/(q^6 w_1))$ and $m_1$ the degree of $\Psi$ in $w_1$.

The construction of the abelian subgroup of $\hat W$ is similar to the cases $B$ and $C$, and is skipped for the sake of brevity.

The minimal degree polynomial solution to the level one $D_r$ $q$KZ system has 
total degree $r(r-1)/2$ and partial degree $m_1=m_r=r-1$ in all variables.
Its base entry, corresponding to the open link pattern $\pi_0$ with only
open arches reads
\begin{equation}
\Psi_{\pi_0}=\prod_{1\leq i<j\leq 2n+1} (qz_i-q^{-1}z_j) 
\end{equation}
and all the other entries of $\Psi$ may be obtained in a triangular way from this one.

Example: for $r=3$, we have the following minimal polynomial solution to the level one $D_3$ $q$KZ
system:
\begin{eqnarray*}
&&\Psi_{\epsfig{file=arch0c.eps,width=0.9cm}}
=(qw_1-q^{-1}w_2)(qw_1-q^{-1}w_3)(qw_2-q^{-1}w_3)\\
&&\Psi_{\epsfig{file=arch1c.eps,width=0.9cm}}
=(qw_1-q^{-1}w_2)(q w_1w_3-q^{-1})(q w_2w_3-q^{-1})\\
&&\Psi_{\epsfig{file=arch2c.eps,width=0.9cm}}
=(q^{-2}-q^{2}w_1^2)(qw_2-q^{-1}w_3)(qw_2w_3-q^{-1}) \\
\end{eqnarray*}
which, upon taking the rational limit gives the multidegrees:
\begin{eqnarray*}
&&\Psi_{\epsfig{file=arch0c.eps,width=0.9cm}}
=(A+z_1-z_2)(A+z_1-z_3)(A+z_2-z_3)\\
&&\Psi_{\epsfig{file=arch1c.eps,width=0.9cm}}
=(A+z_1-z_2)(A+z_1+z_3)(A+z_2+z_3)\\
&&\Psi_{\epsfig{file=arch2c.eps,width=0.9cm}}
=2(A+z_1)(A+z_2-z_3)(A+z_2+z_3) \\
\end{eqnarray*}
and the degrees $\Psi=(1,1,2)$ for $A=1$ and $z_i=0$.

\subsection{RS point and HTASM}
At the point $q=e^{2i\pi/3}$, $\Psi$ may be viewed as the Perron--Frobenius eigenvector 
of a transfer matrix, corresponding in the homogeneous limit to the
Hamiltonian 
\begin{equation}
H_D=e_0+\sum_{i=1}^{r-2}e_i +{e_{r-1}+e_{r-1}'\over 2}
\end{equation}
Note that upon the reflection $e_i\to e_{r-i}$, this Hamiltonian is mapped onto $H_C$:
we are dealing with the same algebra, but in different representations.

Going to the RS point $q=e^{2i\pi/3}$ and taking the homogeneous limit $w_i=1$ for all $i$, 
and normalizing $\Psi$ so that its smallest entry is $\Psi_{\pi_0}=1$,
we have found the
\begin{conjecture*}
The sum of entries $\sum_\pi \Psi_\pi$
is the number of Half-Turn Symmetric Alternating Sign Matrices of size $r$,
$A_{HT}(r)$.
\end{conjecture*}
This conjecture also works in
the even case $r=2n$, which may be obtained from the odd
one by taking $z_1=-q^{-2}$, shifting all remaining spectral parameters $w_i\to w_{i-1}$, $i=2,3,...,2n+1$,
and dividing out by $\prod_{1\leq i\leq 2n} (1+z_i)$. 
Note the formulae $A_{HT}(2n)=\det\Big({i+j\choose 2i-j}+{i+j+1\choose 2i-j}\Big)_{0\leq i,j\leq n-1}$
and $A_{HT}(2n+1)=\det\Big({i+j+1\choose 2i-j}+{i+j+2\choose 2i-j+1}\Big)_{0\leq i,j\leq n-1}$.

Introduce as before the left Perron--Frobenius eigenvector $v$ of $H_D$
with coprime positive integer entries.
\begin{conjecture*}
\begin{equation}
\sum_\pi v_\pi \Psi_\pi = A(r)\ .
\end{equation}
\end{conjecture*}

Finally, we also find the
\begin{conjecture*}
The largest entry of $\Psi$ for $D_r$ is the sum of entries for $C_{r-1}$.
\end{conjecture*}

Example: at $r=5$, $\Psi=(1,1,3,4,2,3,1,4,2,4)$, $v=(10,10,17,14,18,17,23,14,18,25)$, 
$\sum_\pi \Psi_\pi=25=A_{HT}(5)$, 
$\sum_\pi v_\pi \Psi_\pi=429=A(5)$, and the maximal entry of $\Psi$ is $4=A(2)^2$, the sum of the components
of the $C_4$ solution.


%
%
%

\bibliographystyle{amsalpha}

\end{document}